\documentclass[10pt, conference,compsocconf]{IEEEtran}

\usepackage[T1]{fontenc}
\usepackage{color}
\usepackage[english]{babel}
\usepackage{array}
\usepackage{prettyref}
\usepackage{booktabs}
\usepackage{url}
\usepackage{multirow}
\usepackage{amsthm}
\usepackage{amsmath}
\usepackage{amssymb}
\usepackage{graphicx}
\usepackage{flushend}
\usepackage{cite}
\ifCLASSOPTIONcompsoc
\usepackage[caption=false,font=normalsize,labelfont=sf,textfont=sf]{subfig}
\else
\usepackage[caption=false,font=footnotesize]{subfig}
\fi
\usepackage{cite}
\usepackage{MnSymbol}

\theoremstyle{plain}
\newtheorem{thm}{\protect\theoremname}
\theoremstyle{plain}
\newtheorem{prop}[thm]{\protect\propositionname}

\providecommand{\propositionname}{Proposition}
\providecommand{\theoremname}{Theorem}

\ifCLASSINFOpdf
\else
\fi
\usepackage[normalem]{ulem}
\usepackage{hyperref}

\begin{document}

%
\title{Query Processing For The Internet-of-Things: Coupling Of Device Energy Consumption And\ Cloud Infrastructure Billing}

\author{\IEEEauthorblockN{Francesco Renna, Joseph Doyle, Yiannis Andreopoulos}
\IEEEauthorblockA{Dept. of Electronic and Electrical Engineering\\
University College London (UCL)\\
London, UK\\
f.renna,j.doyle,i.andreopoulos@ucl.ac.uk}
\and
\IEEEauthorblockN{Vasileios Giotsas}
\IEEEauthorblockA{Dithen\\
\href{http://www.dithen.com}{http://www.dithen.com}\\
London, UK\\
v.giotsas@dithen.com}
}


%


\maketitle

\begin{abstract}
Audio/visual recognition and retrieval applications have recently garnered significant attention within Internet-of-Things (IoT) oriented services, given that video cameras and audio processing chipsets are now ubiquitous even in low-end embedded systems. In the most typical scenario for such services, each device extracts audio/visual features and compacts them into feature descriptors, which comprise media queries. These queries are uploaded to a remote cloud computing service that performs content matching for classification or retrieval applications. Two of the most crucial aspects for such services are: \textit{(i)} controlling the device energy consumption when using the service; \textit{(ii)} reducing the billing cost incurred from the cloud infrastructure provider. In this paper we derive analytic conditions for the optimal coupling between the device energy consumption and the incurred cloud infrastructure billing. Our framework encapsulates: the energy consumption to produce and transmit audio/visual queries, the billing rates of the cloud infrastructure, the number of devices concurrently connected to the same cloud server, and the statistics of the query data production volume per device. Our analytic results are validated via a deployment with: \textit{(i)} the device side comprising compact image descriptors (queries) computed on Beaglebone Linux embedded platforms and transmitted to Amazon Web Services (AWS) Simple Storage Service; \textit{(ii)} the cloud side carrying out image similarity detection via \ AWS\ Elastic Compute Cloud (EC2) spot instances, with the AWS Auto Scaling being used to control the number of instances according to the demand. 
\end{abstract}


%
\IEEEpeerreviewmaketitle

\section{Introduction}
Most of the envisaged applications and services
for wearable sensors, smartphones, tablets or portable computers in
the next ten years will involve analysis of audio/visual streams for event,
action, object or user recognition, recommendation services and context awareness, etc. \cite{5754008,6616112,6616113,leung2013cloud,soyata2012cloud,girod2011mobile,serra2010audio}.
Examples of early commercial services in this domain include Google
Goggles, Google Glass object recognition, Facebook automatic face
tagging \cite{becker2008evaluation}, Microsoft's Photo Gallery face
recognition, as well as technology described in recent publications
from Google, Siemens and others%
\footnote{See ``A Google Glass app knows what you're looking at'' MIT Tech.
Review (Sept. 30, 2013) and EU projects SecurePhone \cite{sellahewa2005wavelet,bredin2006detecting}
and MoBio \cite{poh2010evaluation,Marcel_CVPR_2010}. %
}. 

\begin{figure}
\begin{centering}
\includegraphics[scale=0.14]{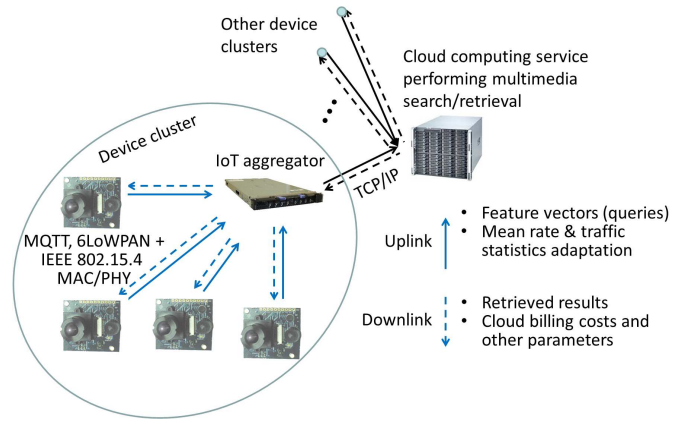} 
\par\end{centering}

\caption{System hierarchy for a media search application within an IoT context. Low-power devices send query data to an IoT aggregator using low-power protocols for the application, network, medium access control and physical layers, such as MQTT, 6LowPAN, and IEEE 802.15.4 MAC/PHY. The IoT aggregator sends aggregated query volumes to the cloud-computing service using TCP/IP. \label{fig:visual-search-system-diagram}}
\end{figure}

Figure \ref{fig:visual-search-system-diagram} presents an example
of how such applications can be deployed in practice within an Internet-of-Things (IoT) context. Energy-constrained devices capture and extract audio/visual features from audio and/or image streams and compact such features into feature-descriptor vectors \cite{arandjelovic2012three,perronnin2010large,jegou2012aggregating,serra2010audio}. Such feature vectors can be seen as \textit{queries} in a multimedia search application \cite{girod2011mobile,arandjelovic2012three}. For example, Serra \textit{et. al.} \cite{serra2010audio} propose beat and tempo feature extraction for cover song identification. A similar service is now widely deployed by Shazam. In the visual search domain,  several approaches produce image salient points and then compact their associated features into compact vectors of 64\(\sim\)8192 elements \cite{jegou2012aggregating,perronnin2010large}. All such compacted feature vectors can be matched to equivalent vectors of very large content libraries within a cloud-computing service within the context of classification, retrieval and similarity identification for so-called ``big data'' applications. Because devices of the same type run the same application software for the query extraction and transmission, they incur, on average, the same energy  consumption per bit of each type of query. Therefore, they can be partitioned into ``device clusters'' that represent a multitude of identical devices (Fig. \ref{fig:visual-search-system-diagram}). An IoT aggregator can be used to aggregate traffic from each device cluster and upload it to a remote cloud computing service that carries out the search operations that provides for recognition
and retrieval purposes \cite{5754008,6616112,6616113,ren2014dynamic}.

In this paper, we consider the energy consumption and billing costs incurred by such applications in a holistic, system-oriented, manner. Specifically, we derive a parametric model that allows for the coupling of the energy consumption and cloud billing costs
in function of the system settings, under the assumption of identical devices producing data traffic with the same statistical characterization.
A key aspect of our model is the derivation of the  \textit{optimal balancing} between:
\begin{enumerate}
\item
 \textit{idle time}, where device energy consumption or cloud billing cost is incurred for no useful output (e.g., image acquisition and processing or buffering on each device that does not lead to query generation, or cloud servers idling due to small volumes of queries being submitted);  
\item
\textit{active time}, where, despite resource consumption being incurred for useful output, one does not want to exceed certain limits in order not to cause excessive energy consumption in the device or excessive billing costs from the cloud infrastructure provider. 
\end{enumerate}
 A key advantage of our work in comparison to previous
work on optimal energy management policies \cite{beninidynamicpowermanagement,ZhangEnergyManagementspringer,alippienergymanagement,ren2014dynamic} and resource prediction and analysis \cite{kontorinis2009statistical,andreopoulos2001local,andreopoulos2003high,andreopoulos2007adaptive}  (see also \cite{Abolfazli14} and references therein), 
is that we provide closed-form expressions for the minimum-required
billing cost in order for each mobile device to remain within the predetermined energy consumption constraints.
In order to validate our analytic derivations, we utilize a proof-of-concept image similarity identification application, deployed via: \textit{(i)} running the image feature extraction and query generation and transmission on a Beaglebone Linux embedded platform; \textit{(ii)} implementing the back-end query processing for similarity identification and retrieval on Amazon Web Services Elastic Compute Cloud (AWS EC2) spot instances. Our results illustrate how the proposed model can be applied to real-world IoT-oriented query retrieval systems in order to establish the desired operational parameters with respect to energy consumption and cloud infrastructure billing. More broadly, the experimental results reported in this paper exemplify the efficacy of our framework for
feasibility studies on energy consumption and billing cost provisioning in cloud-based IoT query processing applications prior to time-consuming testing and deployment. 

The remainder of the paper is organized as follows. In \prettyref{sec:System Model}, we present the system model corresponding
to the application scenarios under consideration. The analytic derivations
characterizing energy-constrained feature extraction are presented in \prettyref{sec:Characterization-of-Energy},
where we also derive the optimal coupling with the utilized cloud-computing service
under three widely-used statistical characterizations for the query production rate. \prettyref{sec:Applications} presents
experimental results and \prettyref{sec:Conclusions} concludes
the paper.

\section{System Model\label{sec:System Model}}

Within the system hierarchy of Fig. \ref{fig:visual-search-system-diagram}, each  mobile device connects to a ``repository''   service of a cloud provider, which represents the collecting unit, i.e. a cloud storage service like AWS Simple Storage Service (S3) or IBM\ IoT Foundation. This is where all device queries are uploaded (e.g., using an application-layer protocol like MQTT) in order to be processed by the back-end search mechanism of the service. As shown in Fig. \ref{fig:visual-search-system-diagram}, an IoT aggregator can be present in-between IoT clusters of the same type and the cloud repository, in order to reshape the IoT query traffic volume before uploading it to the cloud-computing service and also carry out other device-specific and service-specific operations\footnote{Depending on the exact application, the IoT aggregator may carry out  authentication or encryption of queries, reformatting of the retrieved results from the cloud service so that they display correctly on the particular devices, application/collection of device metadata for service statistics and advertising, etc. We do not discuss these aspects as they are out of the scope of the present paper. }. The figure shows that
the essentials of the problem boil down to the analysis of the interaction
between each mobile device node and its corresponding IoT aggregator and cloud computing service.

\subsection{System Description}

We assume that the mobile application is running continuously for a ``monitoring''\ interval of $T$
seconds. This interval corresponds to the typical device usage per day, or in-between battery recharging periods, e.g., $T \in \left[ 60,18000 \right]$ seconds per day. The activation, processing and transmission is either triggered by the user, or by external events at irregular times throughout the application's running time $T$. Examples are: user-triggered audio or visual feature extraction by recording a particular content segment (e.g., as in the Shazam, Google Voice or Google Goggles services), or event-driven activation within an audio/visual surveillance application. We therefore assume that the query data production volume during $T$ seconds  is modeled as a random variable. Finally, we remark that the query data production and transmission and the cloud billing are not strictly continuous processes. However, given that we are focusing on large monitoring intervals comprising tens or hundreds of seconds, they can be seen as continuous processes.

\subsection{Definitions}


\subsubsection{Query Data Production}

The query data production and transmission by each device is
a non-deterministic process, because it depends on the frequency of the application invocation (or on event-driven activation alerts), as well as on the query size, which may vary, depending on the media search application. Therefore, the query data volume (in bits) for each time interval of $T$ seconds
of each device is modeled by random variable (RV) $\Psi_\mathrm{e}$ with probability density function (PDF) $P\left(\psi_\mathrm{e}\right)$.
A model for the marginal statistics of this data volume can be derived by observing the
occurred processing and analyzing the behavior of each device
when it captures image or audio data and produces query bits to be transmitted to the IoT\ aggregator. Examples of systems with variable
query data production and transmission rates include visual sensor networks transmitting
image features \cite{kulkarni2005senseye,rowe2007firefly,tagliasacchiWynerZivGaussian,redondilow},
as well as activity recognition networks where the
data acquisition is irregular and depends on the events occurring
in the monitored area \cite{wernerallen2006fidelity,palma2010distributed,redondi2010laura}.

Beyond individual devices, the query volume uploaded from each IoT aggregator to  the cloud service is modelled by random variable  $\Psi_\mathrm{b}$ with PDF $P\left(\psi_\mathrm{b}\right)$. The distributions  $P\left(\psi_\mathrm{e}\right)$ and  $P\left(\psi_\mathrm{b}\right)$ will be of the same type (the latter will be a scaled version of the former) if the IoT\ aggregator shapes its uploaded traffic in the manner it receives it. Alternatively, if no traffic shaping is performed and the processing latency at the aggregator is fixed,
for an aggregator of $n$ devices:
\begin{equation}
P\left(\psi_\mathrm{b}\right)=\underset{n\:\mathrm{times}}{\underbrace{P\left(\psi_\mathrm{e}\right)\star\ldots\star P\left(\psi_\mathrm{e}\right)}}, \label{eq:P_b_convolution_P_e}
\end{equation}
where $\star$ denotes the convolution operator, i.e., the PDF\ characterizing the uploaded traffic is the result of simple addition of the RVs modelling the data volumes received by all $n$ devices. Note that, as the number of devices $n$ increases, the corresponding PDF $P\left(\psi_\mathrm{b}\right)$ can be approximated with increasing precision via the Half-Gaussian distribution.\footnote{The analysis associated to Half-Gaussian distributed query volumes will be carried out in future works.} Since the query data production volume may
be non-stationary, we assume its marginal statistics for $P\left(\psi_\mathrm{e}\right)$ and  $P\left(\psi_\mathrm{b}\right)$, which are derived starting from a doubly
stochastic model for these processes as explained in related work  \cite{LamGoodmanDCT,foo2008analytical}.

\subsubsection{Energy and Cloud Infrastructure Billing Parameters}

We assume that the production and transmission of one query bit incurs energy consumption rate of $g_\mathrm{e}$ Joule-per-bit (J/b). This rate incorporates the audio or visual capturing, the feature extraction and compaction process to produce the compacted feature vector, and the transmission of the feature vector to the IoT\ aggregator. Since we are considering prolonged periods of operation in our analysis and the utilized sensors, transceivers and embedded processors consume energy in a stable manner when handing data, $g_\mathrm{e}$ can be calculated by averaging several ``on'' periods for sensing, processing and transmission for each device under consideration and normalizing to the amount of bits delivered to the IoT aggregator. For example, under a visual search application, $g_\mathrm{e}$ would incorporate the energy consumption for the image acquisition, the processing to extract salient point descriptions, the compaction process to produce a 256-element feature vector comprising 32-bit numbers  (visual query) corresponding to each image \cite{jegou2012aggregating}, and the application and transceiver-incurred energy consumption to transmit this 8192-bit stream to the aggregator  (e.g., using MQTT\ and the IEEE 802.15.4e MAC/PHY). Assuming that the entire process requires on average $10^{-5}$ J on the mobile device under consideration, this leads to $g_\mathrm{e} \cong 1.2 \times10^{-9}$ J/b. 

On the other hand, given the time-varying nature of the query data production per device,  we also encounter the case where the device is consuming energy to run the application (and possibly capture images or audio) in the background without producing any queries. This corresponds to ``idle'' energy consumption by each device with rate $i_{\mathrm{e}}$ Joule-per-bit (i.e., $i_\mathrm{e}$ Joule for the time interval corresponding to the production and transmission of one query bit). We assume that the application goes in idle mode during time intervals where the amount of produced query bits is below $c_\mathrm{e}E\left[\Psi_\mathrm{e}\right]$ bits, with $E[\Psi_\mathrm{e}]$ the statistical expectation of  $\Psi_\mathrm{e}$. The value of $c_\mathrm{e}$ depends on the processing and transmission capabilities of the device \cite{andreopoulos2007adaptive,kontorinis2009statistical}, as well as on the specifics of the application \cite{andreopoulos2000hybrid,andreopoulos2002new,munteanu2003control}, e.g., the size of the feature vector per query, the manner in which query generation is activated, etc. For instance, regular query generation (e.g., once per second) will correspond to lower value of $c_\mathrm{e}$ in comparison to motion-activated query generation, as the motion detection requires continuous capturing and processing of data that corresponds to higher percentage of ``idle'' energy consumption, i.e., energy consumption that does not lead to query data generation.     

In order to control the overall energy consumption profile of the application, the expected energy consumption within $T$ seconds should be equal to $E_\mathrm{mean}$ Joule and the expected upper-sided deviation should not exceed $E_\mathrm{updev}$ Joule. Both of these values are provided by the application or device developer in order to ensure the application does not degrade the user quality-of-experience (e.g., sudden drop of battery life or device/battery overheating), or disrupt other concurrently-running services on the device.  

Analogously, when servers are reserved from the cloud provider in order to process the queries uploaded by an IoT aggregator, this incurs billing costs. All cloud computing services today use some form of autoscaling mechanism in order to adjust the number of compute instances according to the demand. For example, in AWS Auto Scaling \cite{ryan2015aws} one can set rules that scale the utilized compute instances for every monitoring interval according to the average query volume received during the previous monitoring interval. A typical AWS Auto Scaling setup would be\footnote{The reported numbers of instances and instance types are only indicative and can be adjusted per IoT application.}: 
\begin{itemize}
\item
3 single-core AWS EC2 \texttt{m3.medium} spot instances when the average uploaded query volume is below a certain ``quota'' of $c_\mathrm{b}$ query bits (``idle'' case),  

\item 30 spot instances when the query volume exceeds $c_\mathrm{b}$ bits (``active'' case). 
\end{itemize}
Based on current AWS EC2 pricing, each single-core  \texttt{m3.medium}  spot instance incurs infrastructure billing cost of 0.01\$ per hour. Assuming that a search operation with a $256\times 32$-bit query requires 10ms of cloud service time and under the AWS Auto Scaling rules stated above, this corresponds to billing cost of (approximately): $8.3\times10^{-8}$ dollars-per-query under the ``idle'' case, or $i_\mathrm{b} \cong 1.0 \times 10^{-11} $ dollars-per-query-bit (\$/b);   $p_\mathrm{b}\cong 1.0 \times 10^{-10} $ \$/b for the ``active'' case. The quota of $c_\mathrm{b}$ query bits can be set according to the application or the number of devices, $n$, within each IoT aggregator.  

Beyond the cost of the computing time, billing cost proportional to the expected query volume per monitoring interval, $E[\Psi_\mathrm{b}]$, must be accounted for, since all cloud providers charge for data transfers and storage. Assuming 0.15\$\  per gigabyte of query volume (based on current AWS pricing), this leads to (approximately) $g_\mathrm{b}=1.9\times 10^{-11}$ \$/b. \ Finally, in order to remain competitive against other solutions in the market, the service may wish to set an expectation that each user should be billed for $B_\mathrm{mean}$ \$ on average for each device and each monitoring time interval of $T$ seconds. 

Evidently, the large number of system, data production,  energy consumption, and cloud billing parameters makes the exhaustive exploration of the complete design space infeasible. Therefore, although not all parameters describing the overall system are controlled by the same entity, the creation of an analytic model that can establish closed-form relationships between the different parameters, as well as optimal settings under specified conditions for device energy consumption and billing cost is of the utmost importance. This is the aim of the next section.

\section{Characterization of Energy Consumption and Cloud Billing Cost\label{sec:Characterization-of-Energy}}

We derive analytic expressions for the expected energy consumption of a device (and its upper-side deviate), as well as the expected cloud billing\ for a group of $n$ devices on the same IoT\ aggregator. This allows us to derive closed-form conditions that ensure the value of $E_\mathrm{mean}$ Joule is met for each device, while also meeting the corresponding energy upper-side deviation of $E_\mathrm{updev}$ Joule. We also derive the conditions that minimize the incurred billing cost and ensure that the minimum value can be set to the expected billing of $B_\mathrm{mean}$ per monitoring period of $T$ seconds.  

The expected energy consumption of each mobile device over the monitoring period of $T$ seconds is: 
\begin{equation}
E_{\mathrm{exp}}  =  E\left[\Psi_\mathrm{e}\right]g_{\mathrm{e}}  +  i_\mathrm{e}\int_{0}^{c_\mathrm{e} E\left[\Psi_\mathrm{e}\right] }\left(c_\mathrm{e} E\left[\Psi_\mathrm{e}\right]-\psi_\mathrm{e}\right)P_\mathrm{e}\left(\psi_\mathrm{e}\right)d\psi_\mathrm{e}, \label{eq:E_exp}
\end{equation}
where the integral of the second term expresses the expected energy consumption for the time that the device will be in idle mode. We can also express the one-sided variability of the energy consumption when the application switches from idle to active state (i.e., when exceeding the $(c_\mathrm{e} E\left[\Psi_\mathrm{e}\right])$-bit query volume):
\begin{equation}
E_{\mathrm{var}}  =  g_{\mathrm{e}}^2\int_{c_{\mathrm{e}}E\left[\Psi_\mathrm{e}\right]}^{\infty}\left(\psi_\mathrm{e} - c_\mathrm{e}E\left[\Psi_\mathrm{e}\right]\right)^2 P\left(\psi_\mathrm{e}\right)d\psi_\mathrm{e}. 
\label{eq:E_var}
\end{equation}
Under a given energy budget of $E_\mathrm{exp}$ Joule for the monitoring time interval of $T$ seconds, allowing for a large value for $E_\mathrm{var}$ will incur significant drop in the device battery level (and possibly other unintended consequences, such as device overheating, battery degradation, etc.).  On the other hand, a small value of $E_\mathrm{var}$ will limit the query production volume handled by the device, or may require a very high value for $c_\mathrm{e}$ that may not be realistic for the utilized hardware. Therefore, in this paper we explore this tradeoff by imposing operational values for  the mean energy 

\begin{equation}
E_\mathrm{exp}=E_\mathrm{mean} \label{eq:E_exp_equal_E_mean}
\end{equation}
and the corresponding upper-side energy deviation
\begin{equation}
E_\mathrm{var}= E_\mathrm{updev}^2, \label{eq:E_var_equal_E_updev2}
\end{equation}
and we explore their impact on the system parameters and the query data production volume.

In a similar fashion, let us now consider the expected cloud billing cost when receiving $n$ aggregated query volumes from $n$ devices. We can express this cost via 
\begin{eqnarray}
B_{\mathrm{exp}} & = &    E\left[\Psi_\mathrm{b}\right]g_\mathrm{b} + i_\mathrm{b}\int_{0}^{c_\mathrm{b}}\left(c_\mathrm{b}-\psi_\mathrm{b}\right)P\left(\psi_\mathrm{b}\right)d\psi_\mathrm{b}\nonumber \\ 
& + &  p_\mathrm{b}\int_{c_\mathrm{b}}^{\infty}\left(\psi_\mathrm{b} - c_\mathrm{b}\right)P\left(\psi_\mathrm{b}\right)d\psi_\mathrm{b}, \label{eq:B_exp definition}
\end{eqnarray}
where: $E\left[\Psi_\mathrm{b}\right]g_\mathrm{b}$ corresponds to the data transfer/storage costs, the first integral corresponds to the partial moment expressing the ``idle'' billing cost, and the second integral corresponds to the ``active'' billing.  

Adding and subtracting $p_{\mathrm{b}}\int_{0}^{c_\mathrm{b}}\left(\psi_\mathrm{b}-c_{\mathrm{b}}\right)P\left(\psi_\mathrm{b}\right)d\psi_\mathrm{b}$
in $B_{\mathrm{exp}}$, we get: 
\begin{eqnarray}
B_{\mathrm{exp}} & = & E\left[\Psi_\mathrm{b}\right]\left(g_\mathrm{b} + p_\mathrm{b} \right)-p_\mathrm{b}c_\mathrm{b} \nonumber \\ 
& + &  \left(i_\mathrm{b}+p_\mathrm{b} \right)\int_{0}^{c_\mathrm{b}}\left(c_\mathrm{b}-\psi_\mathrm{b}\right)P\left(\psi_\mathrm{b}\right)d\psi_\mathrm{b}. 
\label{eq:B_exp simplified}
\end{eqnarray}

Evidently, the expected billing cost depends on the coupling point, $c_\mathrm{b}$,
as well as on the PDF of the aggregate query data reaching the cloud service,
$P\left(\psi_\mathrm{b}\right)$, which is either of the same form as $P\left(\psi_\mathrm{e}\right)$, or it is linked to it via \eqref{eq:P_b_convolution_P_e}. In the remainder of this section: 
\begin{itemize}
\item
We consider different cases for $P\left(\psi_\mathrm{e}\right)$ and $P\left(\psi_\mathrm{b}\right)$ to derive
the conditions to match the energy consumption expression of \eqref{eq:E_exp} 
to $E_\mathrm{mean}$ in \eqref{eq:E_exp_equal_E_mean} and allow parameter tuning that guarantees that \eqref{eq:E_var} does not exceed the threshold $E_\mathrm{updev}$ in \eqref{eq:E_var_equal_E_updev2}. 
\item
We derive the number of query bits (quota), $c_\mathrm{b}$,\ that  minimizes the corresponding billing cost of \eqref{eq:B_exp simplified} under various PDFs $P\left(\psi_\mathrm{b}\right)$. 
\item
In order for the desired energy consumption and billing cost parameters to be met concurrently, we associate the minimum billing cost with the desired value for the expected billing, $B_\mathrm{mean}$, and the device query production volume, $r$, thereby establishing the number of devices, $n$, that should be admitted by each IoT\ aggregator. 
\end{itemize}

\subsection{Illustrative Case: $P\left(\psi_\mathrm{e}\right)$ and $P\left(\psi_\mathrm{b}\right)$ are Uniformly Distributed\label{sub:Uniform Case}}

When no knowledge of the underlying statistics of the query generation
process exists, one can assume that both $P\left(\psi_\mathrm{e}\right)$ and $P\left(\psi_\mathrm{b}\right)$ are
uniform over the intervals $\left[0,2r\right]$ and $\left[0,2rn\right]$, respectively: 
\begin{equation}
P_{\mathrm{U}}\left(\psi_\mathrm{e}\right)=\left\{ \begin{array}{c}
\frac{1}{2r},\\
0,
\end{array}\begin{array}{c}
0\leq\psi_\mathrm{e}\leq2r\\
\mathrm{otherwise}
\end{array},\right.\label{eq:P_U_psi_e_uniform}
\end{equation}
and
\begin{equation}
P_{\mathrm{U}}\left(\psi_\mathrm{b}\right)=\left\{ \begin{array}{c}
\frac{1}{2rn},\\
0,
\end{array}\begin{array}{c}
0\leq\psi_\mathrm{b}\leq2rn\\
\mathrm{otherwise}
\end{array}.\right.\label{eq:P_U_psi_b_uniform}
\end{equation}
This corresponds to the case where the IoT aggregator's upload query volume PDF matches the query generation PDF of \eqref{eq:P_U_psi_e_uniform} and the aggregator merges query volumes of $n$ devices. 

The expected value of $\Psi_\mathrm{e}$ is $E_{\mathrm{U}}\left[\Psi_\mathrm{e}\right]=r$
bits and the expected value of  $\Psi_\mathrm{b}$ is $E_{\mathrm{U}}\left[\Psi_\mathrm{b}\right]=rn$
bits. The cases where $c_{\mathrm{e}}>2$ or $c_{\mathrm{b}}>2rn$ are of no practical relevance, because:\ \textit{(i)} the first inequality means each device would always be in idle mode, or \textit{(ii)} the second inequality means the cloud infrastructure would be constantly overprovisioned. Thus, we are only concerned with the case where:  $0<c_{\mathrm{e}}<2$ and $0<c_{\mathrm{b}}<2rn$. 
\subsubsection{Energy Parameter Tuning to Meet the Settings of \eqref{eq:E_exp_equal_E_mean} and \eqref{eq:E_var_equal_E_updev2}}
Starting from the device energy consumption, by using \eqref{eq:P_U_psi_e_uniform} in \eqref{eq:E_exp}, 
we obtain:
\begin{equation}
E_{\mathrm{exp,U}}  =  \left(g_{\mathrm{e}}  + \frac{ i_\mathrm{e} c^2_{\mathrm{e}}}{4} \right)r   \Leftrightarrow\  r =  \frac{  4 E_{\mathrm{exp,U}}}{4 g_{\mathrm{e}} + i_{\mathrm{e}} c_{\mathrm{e}}^2}  .  
\label{eq:E_exp_uniform}
\end{equation}
In addition, by using \eqref{eq:P_U_psi_e_uniform} in \eqref{eq:E_var},
we obtain: 

\begin{equation}
E_{\mathrm{var,U}}  =  g_{\mathrm{e}}^2 \frac{\left(2-c_\mathrm{e}\right)^3}{6}r^2,
 \label{eq:E_var_uniform}
\end{equation}
and by substituting \eqref{eq:E_exp_uniform} in \eqref{eq:E_var_uniform}, we can express the one-side variability of the energy consumption between idle and active state as a function of the idle threshold $c_{\mathrm{e}}$ as
\begin{equation}
E_{\mathrm{var,U}} = \frac{  8 g_{\mathrm{e}}^2 E_{\mathrm{exp,U}}^2 \left(2-c_\mathrm{e}\right)^3  }{3 (4 g_{\mathrm{e}} + i_{\mathrm{e}} c_{\mathrm{e}}^2)^2}.
\label{eq:E_var_c_e}
\end{equation}

Therefore, by imposing the constraint \eqref{eq:E_exp_equal_E_mean} for $E_\mathrm{exp,U}$, we can derive the value of $r$ that matches the expected energy consumption. Moreover, \eqref{eq:E_var_c_e} offers a tool to efficiently tune $c_{\mathrm{e}}$ so that the setting of \eqref{eq:E_var_equal_E_updev2} for $E_\mathrm{var,U}$ is met. In this way, one can carry out a detailed exploration of the mean query production volumes and coupling data volumes per device that satisfy any \textit{a-priori} energy settings for  $E_\mathrm{mean}$ and $E_\mathrm{updev}$, as well as any energy parameters $g_\mathrm{e}$ and $i_\mathrm{e}$, within the monitoring time interval $T$.    

Alternatively, from  \eqref{eq:E_exp_uniform} we can derive the activation threshold $c_{\mathrm{e}}$ that guarantees the average energy consumption, for a given average query volume of $r$ bits, as
\begin{equation}
c_\mathrm{e} = 2 \sqrt{\frac{E_{\mathrm{exp,U}}  - g_{\mathrm{e}}r  }{i_{\mathrm{e}}r}},\label{eq:c_e_Uniform}
\end{equation}
provided that $E_{\mathrm{exp,U}}>g_{\mathrm{e}} r$, which must be the case or else the energy constraint does not suffice for the production of $r$ bits within $T$ seconds.  We also note that the constraint $c_{\mathrm{e}}<2$ implies in this case that $E_{\mathrm{exp,U}}< (g_{\mathrm{e}}+ i_{\mathrm{e}}) r$. These two constraints provide the feasible range for the expected energy consumption under Uniformly-distributed query volumes as: $E_{\mathrm{exp,U}}\in \left( g_{\mathrm{e}} r, (g_{\mathrm{e}}+ i_{\mathrm{e}}) r \right)$.  

Based on \eqref{eq:c_e_Uniform}, the one-side variability of energy consumption can be expressed as a function of the average query volume $r$:
\begin{equation}
E_{\mathrm{var,U}} = \frac{4}{3} g_{\mathrm{e}}^2 r^2 \left(  1 - \sqrt{ \frac{E_{\mathrm{exp,U}} - g_{\mathrm{e}}r   }{i_{\mathrm{e}} r}  }   \right)^3.
\label{eq:E_var,U}
\end{equation}
Via \eqref{eq:E_var,U}, we can numerically determine the value of $r$ for which the corresponding one-sided variability of the energy consumption agrees with the setting of \eqref{eq:E_var_equal_E_updev2}.

\subsubsection{Billing Parameter Tuning to Minimize the Cloud Infrastructure Billing Cost and Meet the Expected Billing $B_\mathrm{mean}$}
We can now turn our attention to the billing cost $B_\mathrm{exp}$ in \eqref{eq:B_exp simplified} for the $n$-device aggregate query production volume over the monitoring time interval of $T$ s. We note that the first and the second derivative of $B_{\mathrm{exp}}$ with respect to the coupling point $c_{\mathrm{b}}$ are given by
\begin{IEEEeqnarray}{rCl}
\frac{d  B_{\mathrm{exp}}}{d c_{\mathrm{b}}}  & = &  - p_{\mathrm{b}}  + (i_{\mathrm{b}}   + p_{\mathrm{b}}) F_{\mathrm{b}} ( c_{\mathrm{b}})  \\
 \frac{d^2  B_{\mathrm{exp}}}{d c_{\mathrm{b}}^2}  & = & (i_{\mathrm{b}}   + p_{\mathrm{b}}) P_{\mathrm{b}} ( c_{\mathrm{b}}) ,
\end{IEEEeqnarray}
where $F_{\mathrm{b}} ( \psi_{\mathrm{b}})$ and $P_{\mathrm{b}} ( \psi_{\mathrm{b}})$ are the cumulative distribution function (CDF) and the PDF of the aggregated query volume $\Psi_{\mathrm{b}}$. Therefore, we can conclude that $B_{\mathrm{exp}}$ is a convex function of $c_{\mathrm{b}}$ when $\Psi_{\mathrm{b}}$ obeys to a continuous distribution with given PDF and CDF. Moreover, the value of $c_{\mathrm{b}}$ that minimizes the billing cost is obtained by solving the equation $\frac{d  B_{\mathrm{exp}}}{d c_{\mathrm{b}}}=0$, i.e.,
\begin{equation}
c_{\mathrm{b}}  = F^{-1} \left(  \frac{p_{\mathrm{b}}}{ i_{\mathrm{b}} + p_{\mathrm{b}} }  \right),
\label{eq:cb_opt_general}
\end{equation}
where $F^{-1} (\cdot)$ represents the inverse function of the CDF of $\Psi_{\mathrm{b}}$. Assuming any strictly-increasing CDF, $c_\mathrm{b}$ will be unique\footnote{Even if the CDF is monotonically increasing, all candidate extrema are equivalent with respect to the derived billing cost.}. Therefore,
in conjunction with the fact that $\forall c_\mathrm{b}: \frac{d^{2}  B_{\mathrm{exp}}}{d c^2_{\mathrm{b}}}>0$, $B_\mathrm{exp}$ has a unique minimum in function of $c_\mathrm{b}$.   

For the case of uniform distribution, by replacing  \eqref{eq:P_U_psi_b_uniform} in \eqref{eq:B_exp simplified}, we obtain the average billing cost as
\begin{equation}
B_{\mathrm{exp,U}}  =  \left(g_\mathrm{b} + p_\mathrm{b} \right)rn-p_\mathrm{b}c_\mathrm{b}  +  \left(i_\mathrm{b}+p_\mathrm{b} \right)\frac{c^2_\mathrm{b}}{4rn}, 
\label{eq:B_exp_uniform}
\end{equation}
and the optimal coupling point is 
\begin{equation}
c_{\mathrm{b,U}}=\frac{ 2p_\mathrm{b}rn}{i_\mathrm{b}+p_\mathrm{b}}.
\label{eq:c_b,U}
\end{equation}
The corresponding minimum-possible billing cost for $c_{\mathrm{b}}\in\left(0,\infty\right)$
is achieved under $c_\mathrm{b}=c_{\mathrm{b,U}}$, and it is:
\begin{equation}
\mathrm{min}\left\{ B_\mathrm{exp,U}\right\}   =  \left(g_\mathrm{b} + p_\mathrm{b} -\frac{p^2_\mathrm{b}}{i_\mathrm{b}+p_\mathrm{b}}\right)rn.
\label{eq:min_B_exp_U}
\end{equation}
The last equation shows that the minimum billing cost increases linearly to the average query data production volume of all $n$ devices. If the minimum value is set to any \textit{a-priori} determined expected billing, i.e., $\mathrm{min}\left\{ B_\mathrm{exp,U}\right\} = B_\mathrm{mean}$, the corresponding device query volume becomes: 
\begin{equation}
r_\mathrm{b,U}   =  \frac{B_\mathrm{mean}}{\left(g_\mathrm{b} + p_\mathrm{b} -\frac{p^2_\mathrm{b}}{i_\mathrm{b}+p_\mathrm{b}}\right)n}.\label{eq:r_b_U}
\end{equation}

\subsubsection{Number of Devices in an IoT Aggregator to Concurrently Satisfy Energy Consumption and Billing Costs}
In order to meet \textit{both} energy and billing costs: $\left\{E_\mathrm{mean},E_\mathrm{updev}\right\} $ and $B_\mathrm{mean}$, we can match the derived query volume of \eqref{eq:r_b_U} with $r_\mathrm{e,U}$ derived from \eqref{eq:E_exp_uniform} and, by tuning $c_{\mathrm{e}}$ via \eqref{eq:c_e_Uniform} and setting $c_\mathrm{b,U}$ to the value given by \eqref{eq:c_b,U}, derive:

\begin{equation}
\ r_\mathrm{b,U}=r_\mathrm{e,U}\Leftrightarrow n_\mathrm{U} =\frac{B_\mathrm{mean}}{\left(g_\mathrm{b} + p_\mathrm{b} -\frac{p^2_\mathrm{b}}{i_\mathrm{b}+p_\mathrm{b}}\right)r_\mathrm{e,U}} . \label{eq:n_U}
\end{equation}
The value of $n_\mathrm{U}$ of \eqref{eq:n_U} comprises the numbers of devices that should be accommodated by an IoT aggregator that receives and transmits queries under the uniform distributions of \eqref{eq:P_U_psi_e_uniform} and \eqref{eq:P_U_psi_b_uniform} when each device meets the energy settings of \eqref{eq:E_exp_equal_E_mean} and \eqref{eq:E_var_equal_E_updev2} and the IoT-uploaded volume leads to minimum billing cost of $B_\mathrm{mean}$.  

Overall, via the energy-constrained analysis that derived  \eqref{eq:E_exp_uniform} and \eqref{eq:E_var_c_e} and the cloud-billing optimization that derived \eqref{eq:c_b,U}--\eqref{eq:n_U}, one can explore different energy and billing settings in order to accommodate: particular types of mobile devices (with given energy consumption parameters), given average query production volume, or given number of devices per IoT cluster of Fig. \ref{fig:visual-search-system-diagram}.

\subsection{Energy-constrained Query Volume Production and Minimum Billing Cost under Pareto and Exponential Distributions\label{sub:Minimum-Expected-Power-OtherPDF}}

We can now extend the previous calculation to other distributions
expressing commonly observed data transmission rates in practical
applications. We consider two additional PDFs for $\Psi$ that have
been used to model the marginal statistics of many real-world data
transmission applications and provide the obtained analytic results
in this subsection. The proofs follow the same process as for
the uniform distribution. For each distribution, we couple its parameters to the average
query volume of the uniform distribution, $r$.
This facilitates comparisons of the energy consumption and billing cost achievable under different
statistical characterizations for the query volume.

\subsubsection{Pareto distribution and fixed query volume}

This distribution has been used, amongst others, to model the marginal
data size distribution of data production processes that result in substantial number
of small data volumes and a few very large ones \cite{paxsonTCP,parkTCP}. 
Consider $P_{\mathrm{P}}\left(\psi_{\mathrm{e}} \right)$ as the Pareto distribution
with scale $v_{\mathrm{e}}$ and shape $\alpha_{\mathrm{e}}>2$ $\left(\alpha_{\mathrm{e}} \in\mathbb{N}\right)$,
\begin{equation}
P_{\mathrm{P}}\left(\psi_\mathrm{e}\right)=\left\{ \begin{array}{c}
\alpha_{\mathrm{e}} \frac{v_{\mathrm{e}}^{\alpha_{\mathrm{e}}}}{\psi_{\mathrm{e}}^{\alpha_{\mathrm{e}}+1}},\\
0,
\end{array}\begin{array}{c}
\psi_\mathrm{e}\geq v_{\mathrm{e}}\\
\mathrm{otherwise}
\end{array}.\right.
\label{eq:P_P_psi_e_Pareto_1}
\end{equation}
The expected value of $\Psi_{\mathrm{e}}$ is $E_{\mathrm{P}}\left[\Psi_{\mathrm{e}}\right]=\frac{\alpha_{\mathrm{e}} v_{\mathrm{e}}}{\alpha_{\mathrm{e}}-1}$
bits. Thus, if we set
\begin{equation}
v_{\mathrm{e}}=\frac{\alpha_{\mathrm{e}}-1}{\alpha_{\mathrm{e}}}r,
\label{eq:v-definition}
\end{equation}
we obtain $E_{\mathrm{P}}\left[\Psi_{\mathrm{e}}\right]=r$ bits, i.e., we
match the expected query volume per device to that of the Uniform distribution. The characterization of the energy consumption for queries with Pareto-distributed volumes is summarized in the following proposition.

\begin{prop}
The average energy consumption for Pareto-distributed device query volumes is given by
\begin{IEEEeqnarray}{rCl}
E_{\mathrm{exp,P}} & =&  \left[  g_{\mathrm{e}}  + i_{\mathrm{e}}  \left[  (\alpha_{\mathrm{e}}-1)^{\alpha_{\mathrm{e}}-1} c_{\mathrm{e}} (\alpha_{\mathrm{e}} c_{\mathrm{e}})^{-\alpha_{\mathrm{e}}} + c_{\mathrm{e}}-1  \right]    \right]r, \IEEEeqnarraynumspace
 \label{eq:E_exp_P_r_1}
\end{IEEEeqnarray}
and the one-sided variation of the energy consumption from idle mode to active mode is given by
\begin{IEEEeqnarray}{rCl}
E_{\mathrm{var,P}} & = & 2 g_{\mathrm{e}}^2 \frac{ (\alpha_{\mathrm{e}}-1)^{\alpha_{\mathrm{e}}-1}c_{\mathrm{e}}^{2 - \alpha_{\mathrm{e}}}  }{\alpha_{\mathrm{e}}^{\alpha_{\mathrm{e}}}(\alpha_{\mathrm{e}}-2)} r^{2}.
\label{eq:E_var_P_r_1}
\end{IEEEeqnarray}

\end{prop}

\begin{IEEEproof}
See Appendix.
\end{IEEEproof}

Note that Proposition 1 assumes that $c_{\mathrm{e}} \geq \frac{\alpha_{\mathrm{e}}-1}{\alpha_{\mathrm{e}}}$, since, otherwise, the device will never switch from active to idle state. Moreover, from \eqref{eq:E_exp_P_r_1}, we can derive the average query volume corresponding to any given values for $E_{\mathrm{exp,P}}$ and $c_\mathrm{e}$ as
\begin{equation}
r = \frac{E_{\mathrm{exp,P}}}{ g_{\mathrm{e}}  + i_{\mathrm{e}}  \left[ (\alpha_{\mathrm{e}}-1)^{\alpha_{\mathrm{e}}-1} c_{\mathrm{e}} (\alpha_{\mathrm{e}} c_{\mathrm{e}})^{-\alpha_{\mathrm{e}}} + c_{\mathrm{e}}-1  \right]   },
\label{eq:r_E_exp_P_1}
\end{equation}
and the one-sided energy variance associated to $r$ as
\begin{IEEEeqnarray}{rCl}
\nonumber
E_{\mathrm{var,P}} & = &  g_{\mathrm{e}}^2 \frac{ (\alpha_{\mathrm{e}}-1)^{\alpha_{\mathrm{e}}-1}c_{\mathrm{e}}^{2 - \alpha_{\mathrm{e}}}  }{\alpha_{\mathrm{e}}^{\alpha_{\mathrm{e}}}(\alpha_{\mathrm{e}}-2)} \\
& & \times \frac{  E_{\mathrm{exp,P}}^{2}  }{  \left[  g_{\mathrm{e}}  + i_{\mathrm{e}}  \left[  (\alpha_{\mathrm{e}}-1)^{\alpha_{\mathrm{e}}-1} c_{\mathrm{e}} (\alpha_{\mathrm{e}} c_{\mathrm{e}})^{-\alpha_{\mathrm{e}}} + c_{\mathrm{e}}-1  \right]    \right]^{2}}. \IEEEeqnarraynumspace
\end{IEEEeqnarray}
A particular case of interest for the Pareto distribution arises when  $\alpha_{\mathrm{e}} \to +\infty$: in this limit case, the query volume per device converges to the expectation $E_{\mathrm{P}}\left[\Psi_{\mathrm{e}}\right]=r$, i.e., to \textit{fixed-volume} query production per monitoring interval. Then, since $c_{\mathrm{e}}\geq \frac{\alpha_{\mathrm{e}}-1}{\alpha_{\mathrm{e}}}$, the average energy consumption converges to
\begin{equation}
E_{\mathrm{exp,P}} = \left[g_{\mathrm{e}} + i_{\mathrm{e}} (c_{\mathrm{e}}-1)\right]r,
\end{equation}
as $\alpha_{\mathrm{e}} \to \infty$, and the one-side energy variation from idle to active mode converges to zero (the device is in idle mode for a portion of the time of every monitoring interval). Then, the average query volume that meets the expected energy consumption constraint $E_{\mathrm{exp,P}}$ is simply given by
\begin{equation}
r = \frac{E_{\mathrm{exp,P}}}{g_{\mathrm{e}} + i_{\mathrm{e}} (c_{\mathrm{e}}-1)}.
\end{equation}

\subsubsection{Exponential distribution}

This distribution is relevant to our application context since the marginal statistics of compressed image and video traffic have often been modeled
as exponentially decaying \cite{daiMPEG4exponential}. Consider $P_{\mathrm{E}}\left(\psi_\mathrm{e}\right)$
as the Exponential distribution with rate parameter $\frac{1}{r}$
\begin{equation}
P_{\mathrm{E}}(\psi_{\mathrm{e}})  = \frac{1}{r} \exp \left(  -\frac{1}{r} \psi_{\mathrm{e}}   \right),
\label{eq:P_P_psi_e_Exponential_1}
\end{equation}
for $\psi_{\mathrm{e}} \geq 0$. In this case, the expected value of $\Psi_{\mathrm{e}}$ is $E_{\mathrm{E}}\left[\Psi_{\mathrm{e}}\right]=r$ bits. The characterization of the energy consumption for queries with exponentially distributed volumes is summarized in the following proposition.

\begin{prop}
The average energy consumption for Exponentially-distributed device query volumes is given by
\begin{equation}
E_{\mathrm{exp,E}} = \left[  g_{\mathrm{e}}  + i_{\mathrm{e}} \left(c_{\mathrm{e}} + e^{-c_{\mathrm{e}} } -1  \right)  \right] r,
\label{eq:E_exp_exponential_1}
\end{equation}
and the one-sided variation of the energy consumption from idle mode to active mode is given by
\begin{equation}
E_{\mathrm{var,E}} = 2 g_{\mathrm{e}}^2 \mathrm{exp}(-c_{\mathrm{e}}) {r^2} .
\label{eq:E_var_exponential_}
\end{equation}
\end{prop}
\begin{IEEEproof}
See Appendix.
\end{IEEEproof}

From \eqref{eq:E_exp_exponential_1}, it is straightforward to derive the average query volume corresponding to any given values of  $E_{\mathrm{exp,E}}$ and $c_{\mathrm{e}}$ as
\begin{equation}
r = \frac{E_{\mathrm{exp,E}}}{g_{\mathrm{e}}  + i_{\mathrm{e}} \left[c_{\mathrm{e}} + \mathrm{exp}({-c_{\mathrm{e}} }) -1  \right] },
\label{eq:r_E_exp_exponential_1}
\end{equation}
and the one-sided energy variation associated to $r$ as
\begin{equation}
E_{\mathrm{var,E}} = \frac{  2 g_{\mathrm{e}}^2 \mathrm{exp}({-c_{\mathrm{e}}})  E_{\mathrm{exp,E}}^2}{\left[  g_{\mathrm{e}}  + i_{\mathrm{e}} \left(c_{\mathrm{e}} + \mathrm{exp}({-c_{\mathrm{e}} }) -1  \right)  \right]^2}.
\end{equation}
In addition, for any given values of $E_{\mathrm{exp,E}}$ and $r$, we can also derive the  threshold $c_\mathrm{e}$ as
\begin{equation}
c_{\mathrm{e}} = W_0 \left( - \exp\left(  -  \frac{  E_{\mathrm{exp,E}} + i_{\mathrm{e}}r - g_{\mathrm{e}}r  }{i_{\mathrm{e}}r}   \right)      \right) + \frac{  E_{\mathrm{exp,E}} + i_{\mathrm{e}}r - g_{\mathrm{e}}r  }{i_{\mathrm{e}}r},
\end{equation}
where $W_0(\cdot)$ is the main branch of the standard Lambert W function. The corresponding one-sided energy variability associated to $c_{\mathrm{e}}$ is given by
\begin{equation}
E_{\mathrm{var,E}}  = - 2 g_{\mathrm{e}}^2 r^2 W_0 \left( - \exp\left(  -  \frac{  E_{\mathrm{exp,E}} + i_{\mathrm{e}}r - g_{\mathrm{e}}r  }{i_{\mathrm{e}}r}   \right)      \right).
\end{equation}

\subsubsection{Billing Cost under Pareto and Exponential Distribution}
We now consider the billing cost for the processing of queries uploaded from $n$  devices via an IoT aggregator.
Let us first consider the aggregate query volume distribution modeled via a Pareto distribution with mean $E_{\mathrm{P}}[\Psi_{\mathrm{b}}]=r n$, i.e., 
\begin{equation}
P_{\mathrm{P}}\left(\psi_\mathrm{b}\right)=\left\{ \begin{array}{c}
\alpha_{\mathrm{b}} \frac{v_{\mathrm{b}}^{\alpha_{\mathrm{b}}}}{\psi_{\mathrm{b}}^{\alpha_{\mathrm{b}}+1}},\\
0,
\end{array}\begin{array}{c}
\psi_\mathrm{b}\geq v_{\mathrm{b}}\\
\mathrm{otherwise}
\end{array}\right. ,
\label{eq:P_P_psi_b_Pareto_1}
\end{equation}
where $\alpha_{\mathrm{b}}>2$ $\left(\alpha_{\mathrm{b}} \in\mathbb{N}\right)$ and $v_{\mathrm{b}}=\frac{\alpha_{\mathrm{b}}-1}{\alpha_{\mathrm{b}}}r n.$
\begin{prop}
\label{prop:B}
The average billing cost incurred from processing Pareto-distributed query volumes is given by
\begin{equation}
B_{\mathrm{exp,P}}   =  
(g_{\mathrm{b}} - i_{\mathrm{b}}) r n + (i_{\mathrm{b}} + p_{\mathrm{b}})  \frac{(\alpha_{\mathrm{b}}-1)^{\alpha_{\mathrm{b}}-1}}{\alpha_{\mathrm{b}}^{\alpha_{\mathrm{b}}}}(r n)^{\alpha_{\mathrm{b}}} c_{\mathrm{b}}^{1-\alpha_{\mathrm{b}}} + i_{\mathrm{b}} c_{\mathrm{b}}.
\end{equation}
The minimum billing cost is obtained when
\begin{equation}
c_{\mathrm{b,P}} =  
\left(   \frac{i_{\mathrm{b}}  + p_{\mathrm{b}}}{i_{\mathrm{b}}}   \right)^{\frac{1}{\alpha_{\mathrm{b}}}} \frac{\alpha_{\mathrm{b}}-1}{\alpha_{\mathrm{b}}} r n,
\label{eq:c_b,P_1}
\end{equation}
and it is given by
\begin{IEEEeqnarray}{rCl}
\min \{  B_{\mathrm{exp,P}}  \} & = & \left[ g_{\mathrm{b}} - i_{\mathrm{b}}   + i_{\mathrm{b}} \left(   \frac{i_{\mathrm{b}}+ p_{\mathrm{b}}}{i_{\mathrm{b}}}  \right)^{\frac{1}{\alpha_{\mathrm{b}}}}  \right] r n.
 \label{eq:minBex}
\end{IEEEeqnarray}
\end{prop}

\begin{IEEEproof}
The proof stems from the evaluation of the general solution expressed in \eqref{eq:cb_opt_general} under the usage of the Pareto PDF.
\end{IEEEproof}

In order to ensure that the average billing cost is $B_{\mathrm{mean}}$ and average query volume per device is $r_{\mathrm{e,P}}$, the IoT aggregator must handle 
\begin{IEEEeqnarray}{rCl}
n_{\mathrm{P}} & = & \frac{B_{\mathrm{mean}}}{r_{\mathrm{e,P}}   \left[ g_{\mathrm{b}} - i_{\mathrm{b}}   + i_{\mathrm{b}} \left(   \frac{i_{\mathrm{b}}+ p_{\mathrm{b}}}{i_{\mathrm{b}}}  \right)^{\frac{1}{\alpha_{\mathrm{b}}}}  \right] }
\end{IEEEeqnarray}
devices. This is derived by setting $\min \{  B_{\mathrm{exp,P}}\}=B_{\mathrm{mean}}$ in \eqref{eq:minBex} and solving for $n$.
We also note that, when assuming that the aggregate query volume is Pareto distributed, by letting $\alpha_{\mathrm{b} } \to +\infty$, we can analyze the case when the aggregate query volume at the IoT is fixed and equal to $rn$. In this case, if $c_{\mathrm{b}} \geq r n$, the average billing cost is simply given by
\begin{equation}
B_{\mathrm{exp,P}} = (g_{\mathrm{b}}-i_{\mathrm{b}}) rn + i_{\mathrm{b}} c_{\mathrm{b}},
\end{equation}
which is minimized by setting $c_{\mathrm{b}}$ equal to the mean, i.e., $c_{\mathrm{b,P}} = r n$.

Finally, let us consider the aggregate query volume distribution modeled via an Exponential distribution with mean $E_{\mathrm{E}}[\Psi_{\mathrm{b}}] = rn$, i.e., 
\begin{equation}
P_{\mathrm{E}}(\psi_{\mathrm{b}})  = \frac{1}{rn} \exp \left(  -\frac{1}{r n} \psi_{\mathrm{b}}   \right),
\label{eq:P_P_psi_b_Exponential_1}
\end{equation}
for $\psi_{\mathrm{b}} \geq 0$.
 \begin{prop}
The average billing cost incurred from processing Exponentially-distributed query volumes is given by
\begin{equation}
B_{\mathrm{exp,E}} = (g_{\mathrm{b}} - i_{\mathrm{b}}) r n + i_{\mathrm{b}}c_{\mathrm{b}} +(i_{\mathrm{b}} + p_{\mathrm{b}}) n r e^{-\frac{c_{\mathrm{b}}}{nr}}.
\end{equation}
The minimum billing cost is obtained when
\begin{equation}
c_{\mathrm{b,E}} = {rn}  \ln \frac{i_{\mathrm{b}} + p_{\mathrm{b}}}{i_{\mathrm{b}}},
\label{eq:c_b,E_1}
\end{equation}
and it is given by
\begin{equation}
\min \{ B_{\mathrm{exp,E}} \} = \left( g_{\mathrm{b}} + i_{\mathrm{b}} \ln \frac{i_{\mathrm{b}} + p_{\mathrm{b}}}{i_{\mathrm{b}}}    \right) r n.
\label{eq:Bmin_E}
\end{equation}
\end{prop}
\begin{IEEEproof}
The proof stems from the evaluation of the general solution expressed in \eqref{eq:cb_opt_general} under the usage of the Exponential PDF.
\end{IEEEproof}

In order to ensure that the average billing cost is $B_{\mathrm{mean}}$ and average query volume per device is $r_{\mathrm{e,P}}$, the IoT aggregator must handle
\begin{IEEEeqnarray}{rCl}
n_{\mathrm{E}} & = & \frac{B_{\mathrm{mean}}}{ r_{\mathrm{e,E}}   \left( g_{\mathrm{b}} + i_{\mathrm{b}} \ln \frac{i_{\mathrm{b}} + p_{\mathrm{b}}}{i_{\mathrm{b}}}    \right)  }
\end{IEEEeqnarray}
devices.  This is derived by setting $\min \{  B_{\mathrm{exp,E}}\}=B_{\mathrm{mean}}$ in \eqref{eq:Bmin_E} and solving for $n$.

\section{Evaluation of the Analytic Results\label{sec:Applications}}

To validate the proposed analytic modeling framework of
Propositions 1--4, we performed a series of experiments based on a visual
sensor network connected to an IoT aggregator, and eventually to an AWS S3 plus EC2 cluster of spot instances. The following subsections present the hardware and application specifications, as well as the achieved results. Beyond our specific experimental results, we ensure to retain our description as broad as possible in order to indicate ways to carry out similar experiments within other IoT-oriented platforms, such as IBM\ IoT\ Foundation and Bluemix, AWS IoT, Cisco OpenStack, etc.  

\subsection{System Specification}
We utilized a visual sensor network composed of multiple BeagleBone Linux embedded platforms \cite{redondi2014energy,CancliniSENSYS2013}. Each
BeagleBone is equipped with a RadiumBoard CameraCape board to provide for
the video frame acquisition. For energy-efficient processing, we downsampled
all input images to QVGA (320x240) resolution. Further, our deployment involved: 
\begin{enumerate} 
\item
a portable computer acting as the IoT aggregator, i.e., collecting all bitstreams via a star topology with $n=10$ nodes and the recently-proposed (and available as
open source) TFDMA protocol \cite{burana2012DTFDMA} for contention-free
MAC-layer coordination; 
\item an AWS\ S3 bucket where the IoT aggregator uploads all queries via a TCP/IP connection using script code running on the AWS\ Command Line Interface; 
\item
One reserved AWS instance running as the control server and assigning query volumes from S3 to AWS\ EC2 spot instances that serve as compute units; via AWS\ Auto Scaling, within each monitoring instance of $T$\ seconds, the number of spot instances are set to: 
\begin{itemize}
\item
3 when the query volume is below $c_\mathrm{b}$ bits (``idle'' case). 
\item
30 when the query volume exceeds $c_\mathrm{b}$ bits (``active'' case).  
\end{itemize}
Under our deployment and the utilized application, the uploaded query vectors are matched with the feature vectors extracted from $80,000$ images of similar content. The corresponding billing rates per query bit for this matching operation were found to be  $i_\mathrm{b}=6.27\times 10^{-11}$ \$/b and $p_\mathrm{b}=6.27\times 10^{-10}$ \$/b. Regarding query traffic upload and storage costs, the corresponding billing rate per query bit was found to be  $g_\mathrm{b}=2.09\times 10^{-10}$ \$/b,
\end{enumerate}
We note that no WiFi or other IEEE802.15.4
networks were concurrently operating in the utilized channels of the 2.4
GHz band. However, even if IEEE 802.11 or other IEEE 802.15.4 networks coexist
with the proposed deployment, well-known channel hopping schemes like
TSCH \cite{TSCH} can be used at the MAC layer to mitigate such external interference. Moreover, experiments have shown that such protocols
can scale to hundreds or even thousands of nodes \cite{TSCH}.
Therefore, our evaluation is pertinent to such scenarios that may
be deployed in the next few years within an IoT paradigm \cite{gubbi2013IoT}.
\subsection{Visual Similarity Identification Based on the Vector of Locally Aggregated Descriptors (VLAD) }
Each BeagleBone runs a basic motion detection algorithm (based on successive frame differencing) that generates a visual query only when sufficient motion is detected between  the captured video frames. The query vectors were generated using the state-of-the-art VLAD algorithm of Jegou \textit{et. al.} \cite{jegou2012aggregating}, which is based on SIFT feature extraction and compaction using local feature centers and a PCA projection matrix, both of which are derived offline via training with representative video data \cite{jegou2012aggregating}. The VLAD descriptor (i.e., query) size was set to 256 coefficients of 32 bits each.  

With respect to the visual feature extraction, dedicated energy-measurement tests were performed with the Beaglebone following the energy measurement setup of our previous work  \cite{redondi2014energy} (repeated tests with a resistor in series to the Beaglebone board and a high-frequency oscilloscope to capture the power consumption profile across repeated monitoring intervals). Under the utilized setup, we measured the average energy cost to produce and transmit a query bit, as well as the average
initialization cost per frame for both application scenarios. The resulting energy rates were: $g_\mathrm{e}=1.78\times 10^{-6}$ J/b and $i_\mathrm{e}=6.10\times 10^{-7}$ J/b. Moreover, under the utilized application, the Beaglebone can process up to 1 frame per second while being constantly active. Therefore, the maximum query rate is 1 query per second, i.e., $8192$ b/s. By setting mean query rates such that $E[\Psi_\mathrm{e}]\leq2048$ bits per second, this theoretically allows for ``idle'' energy consumption with $c_\mathrm{e}<3$. In practice, we only utilized   $c_\mathrm{e}\in(0,2)$ for ``idle'' energy consumption (i.e., up to twice the number of frames captured and processed with no query generation), as higher values caused system instability.

\subsection{Results with Controlled Query Generation that Matches the Marginal PDFs Considered in the Theoretical Analysis}

\label{subsec:model_validation}

Under the settings described previously, our first goal is to validate the analytic expressions of Section \ref{sec:Characterization-of-Energy} that form the mathematical foundation
for Propositions 1--2.
To this end, we create a controlled query data production process
on each node by: \emph{(i)} artificially setting several sets
of query volumes according to the marginal PDFs of Section \ref{sec:Characterization-of-Energy}
via rejection sampling \cite{gilks1992adaptive}, a.k.a., Monte Carlo sampling; \emph{(ii)} setting
the mean query volume size per monitoring interval, $r$, to predetermined values. The
sets containing query volume sizes are preloaded onto the memory of
each sensor node during the setup phase. At run time, each node runs a special routine, which, per monitoring interval $t$: \textit{(i)} reads
the corresponding query volume size, $v(t)$, from the preloaded set; \textit{(ii)} captures and processes  $\frac{v(t)}{8192}$ frames, \textit{(iii)} transmits the produced $v(t)$ query bits to the IoT aggregator; \textit{(iv)} if $v(t)<c_\mathrm{e}E[\Psi_\mathrm{e}]$, captures and processes    $\frac{c_\mathrm{e}E[\Psi_\mathrm{e}]-v(t)}{8192}$ additional frames without transmitting queries. In this way, we emulate the actual operation of the node under various query volumes that match the statistical models considered by our analysis and various thresholds $c_\mathrm{e}$ for switching between ``idle'' and ``active'' states. This controlled experiment is designed to confirm the validity of our analytic derivations when the operating conditions match the model assumptions precisely. 

\begin{figure}
\begin{centering}
\includegraphics[scale=0.45]{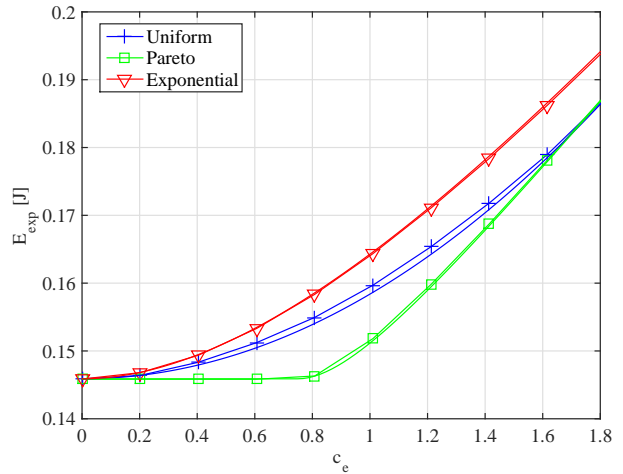} \par\end{centering}
\caption{Average energy consumption $E_{\mathrm{exp}}$ vs. $c_{\mathrm{e}}$. The average query volume was set to $r= 81,920$ b. For the case of Pareto distribution, we used $\alpha_{\mathrm{e}}=4$. Lines with markers: Monte Carlo experiments; Lines without markers: theoretical predictions. 
\label{fig:Eexp_MC}}
\end{figure}

\begin{figure}
\begin{centering}
\includegraphics[scale=0.45]{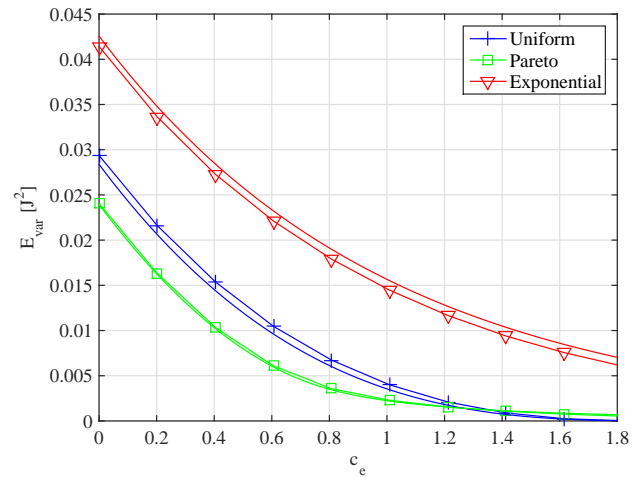} 
\par\end{centering}
\caption{One-sided energy consumption $E_{\mathrm{var}}$ vs. $c_{\mathrm{e}}$. The average query volume was set to $r= 81,920$ b. For the case of Pareto distribution, we used $\alpha_{\mathrm{e}}=4$. Lines with markers: Monte Carlo experiments; Lines without markers: theoretical predictions.
\label{fig:Evar_MC}}
\end{figure}

Indicative experimental results for monitoring time interval of $T=60$ s are reported in  Fig. \ref{fig:Eexp_MC} and Fig. \ref{fig:Evar_MC} for $r=81,920$ b. It is evident that
the theoretical results match the Monte Carlo experiments regarding energy consumption for all the
tested distributions, with all the  $R^{2}$
values (coefficients of determination) between the experimental and the model points being above $0.998$.
We have observed the same level of accuracy
for the proposed model under a variety of data sizes ($r$) and active
time interval durations ($T$), but
omit these repetitive experiments for brevity of exposition.

Similar experiments have been carried out in order to validate the analytic expressions of Propositions 3 and 4 regarding the average billing cost. Specifically, we have submitted indicative queries to the cloud-computing service with volumes that have been generated according to the marginal PDFs of Section \ref{sec:Characterization-of-Energy}
via rejection sampling under various numbers of devices per IoT cluster ($n$) and various average query volumes. The aggregated queries are uploaded to the dedicated S3 bucket for the service and are processed by a number of instances that is controlled by the AWS Auto Scaling rules stated in the previous subsection. In this case, we used $T=600$ s and varied the value of $c_\mathrm{b}$ in order to see the incurred infrastructure billing costs under a variety of Auto Scaling thresholds.

\begin{figure}
\begin{centering}
\includegraphics[scale=0.45]{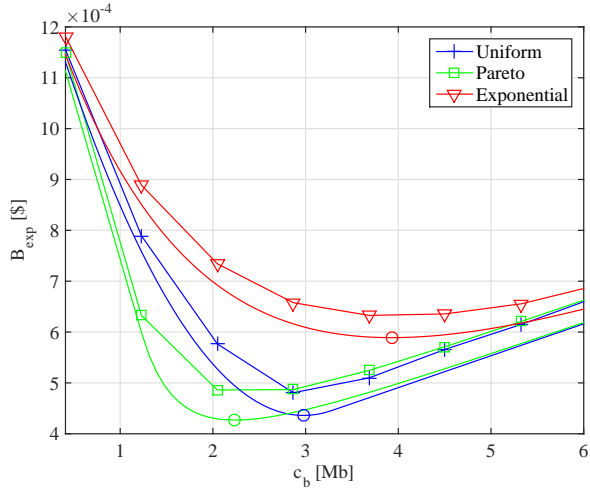} 
\par\end{centering}
\caption{Average billing cost $B_{\mathrm{exp}}$ vs. $c_{\mathrm{b}}$. The average query volume per device was set to $r= 163,840$ b and the experiment corresponds to $n=10$ devices. For the case of Pareto distribution, we used $\alpha_{\mathrm{e}}=4$. Lines with markers: Monte Carlo experiments; Lines without markers: theoretical predictions. The circles indicate minimum billing values as predicted by the analysis in Section \ref{sec:Characterization-of-Energy}.
 \label{fig:Bexp_MC}}
\end{figure}

Fig.~\ref{fig:Bexp_MC} presents indicative results under this setup. Evidently, the theoretical results follow the trends of the experimental data, with  $R^2$ coefficients being above $0.9983$ for all the distributions under consideration. However, the theoretical predictions always tend to underestimate the experimental values. This underestimation is due to the fact that our analysis does not take into account some practical latency and cost aspects of the service, for example that switching between ``idle'', ``active'' states is not instantaneous and other cost overheads (such as the cost of the control server) are not taken into account by our analysis. Similar results to Fig. \ref{fig:Bexp_MC} have been obtained for a variety of average query volumes and monitoring intervals, but are omitted for brevity of exposition.

\subsection{Results with User Generated Data}


We now present system tuning results when repeating the visual query generation, transmission and cloud-based processing for 25 monitoring intervals based on real video captures and VLAD query generation using real data. The experiment was carried out within several offices of the Electronic and Electrical Engineering Department of University College London, and activation of query generation, transmission and processing was triggered when people passed (or moved) in front of the device cameras. Back-end query similarity identification was done using prestored VLAD signatures of $80,000$ images of similar content based on the AWS setup described in the previous subsection.  

\begin{figure}
\begin{centering}
\includegraphics[scale=0.45]{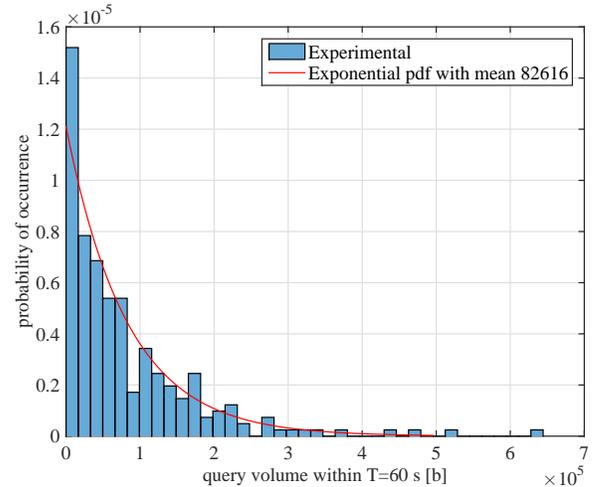} 
\par\end{centering}
\caption{Probability histogram of query volume for $T=60$ s and the best fit obtained via the Exponential distribution.  \label{fig:fit-exponential-pdf}}
\end{figure}

Once data has been collected, we fitted%
\footnote{Fitting is performed by matching the average data size $r$ of each
distribution to the average data size of the JPEG compressed frames
or the set of visual features.%
} the query production volumes to one of the distributions used
in Section \ref{sec:Characterization-of-Energy}. In the performed experiment and under monitoring interval of $T=60$ s for the devices, we found
that the query volume histogram agreed best with the Exponential distribution with  $r=82,616$ b. For $T=\{600,1200\}$ s, the best fit was found to be the Pareto distribution with: $r=816,250$ b and $\alpha=3.89$, and  $r=1,569,700$ b and $\alpha=3.95$, respectively. An example for the fit obtained with the Exponential distribution is given in Fig. \ref{fig:fit-exponential-pdf}. Moreover, with respect to $c_\mathrm{e}$, we found that, for all cases of monitoring intervals under consideration, the system switched between ``idle'' and ``active'' states at $c_\mathrm{e}\cong0.5$. Therefore, our analytic results utilized this value for all results of this subsection.

Under this setup and with the fitted values for Exponential and Pareto PDFs, Table \ref{tab:energy-results} presents the obtained experimental and theoretical values (via Proposions 1 and 2)\ for the expected energy and the upper-sided energy variance for two monitoring intervals.  It is observed that the theoretical predictions are always within 10\% of\ the experimentally-derived values. As such, the proposed energy-consumption model
can be used for early-stage testing of plausible application
deployments with respect to their energy efficiency in order to determine
the impact of various options, prior
to more detailed experimentation in the field. 

To present a further example of this capability, with the Pareto-distributed query volume statistics for $T=\{600,1200\}$ s and under the use of $n=10$ devices, we determined the Auto Scaling threshold, $c_\mathrm{b}$, that is expected to lead to the minimum cloud infrastructure billing cost based on Proposition 3. Then, we benchmarked the obtained cost of the system under this  threshold against the intuitive (albeit \textit{ad-hoc}) setting of $c_\mathrm{b}=nr$, which corresponds to the Auto Scaling threshold being set to match the average query volume of all $n$ devices. The results, given in Table \ref{table:B}, show that the obtained billing cost is 19\% lower than the case of the same query volume processing under the ad-hoc Auto Scaling threshold. In terms of practical deployments, it is important to emphasize again that not all the system parameters can be tuned by the same entity. For example, $c_\mathrm{b}$ is controlled by the cloud provider, whereas $c_{\mathrm{e}}$ depends on the specific device and the processing task performed. However, the proposed framework provides an analytic link between such parameters and the energy and billing costs, that can be used by the different stakeholders in a variety of ways. Moreover, the experimental example reported demonstrates that tuning the system based on the theoretical analysis can lead to important cost savings under real-world conditions for cloud-based processing of IoT-generated queries.

\begin{table}
\caption{\label{tab:energy-results}Expected energy consumption and upper-sided variation. Experimental results and theoretical prediction. For all cases, we set $c_{\mathrm{e}}=0.5$.}
\begin{center}
\begin{tabular}{r|m{22mm}m{22mm}}
 &Theoretical& Experimental \\
 \toprule
  $T=60$ s  & $E_{\mathrm{exp}} = 0.1679 $ J $E_{\mathrm{var}} = 0.0316$ J$^2$& $E_{\mathrm{exp}} = 0.1538$ J $E_{\mathrm{var}} = 0.0317$ J$^2$ \\
 \midrule
$T = 1200$ s &   $E_{\mathrm{exp}} = 2.7955 $ J $E_{\mathrm{var}} = 2.9276$ J$^2$ & $E_{\mathrm{exp}} = 2.8053$ J $E_{\mathrm{var}} = 2.8411$ J$^2$ \\
\bottomrule
\end{tabular}
\label{table:E}
\end{center}
\end{table}

\begin{table}
\caption{\label{tab:billing-results}Expected billing cost. The ad-hoc solution corresponds to setting $c_{\mathrm{b}} = r n$. The proposed solution is obtained with $c_{\mathrm{b}}$ set according to Proposition~\ref{prop:B}.}
\begin{center}
\begin{tabular}{m{14mm}|m{25mm}m{25mm}|m{8mm}}
  & Ad-hoc & Proposition 3 & Saving \\
 \toprule
 $T=600$ s $n=10$  &    $B_{\mathrm{exp}} = 5.82 \cdot 10^{-4}$ \$ $c_\mathrm{b}=1.54$ Mb & $B_{\mathrm{exp}} = 4.70 \cdot 10^{-4}  $ \$ $c_\mathrm{b}=2.51$ Mb  & 19 \% \\
 \midrule
$T = 1200$ s $n=10$   & $B_{\mathrm{exp}} = 7.70 \cdot 10^{-4} $ \$  $c_\mathrm{b}=1.88$ Mb& $ B_{\mathrm{exp}} =6.25 \cdot 10^{-4}  $ \$  $c_\mathrm{b}=3.09$ Mb & 19 \%  \\
\bottomrule
\end{tabular}
\label{table:B}
\end{center}
\end{table}

\section{Conclusions\label{sec:Conclusions}}
We propose a novel theoretical framework for establishing trade-offs in the energy consumption and infrastructure billing cost of Internet-of-Things (IoT)\ oriented deployments comprising mobile devices generating queries that are processed by a back-end cloud computing service. Our analysis incorporates energy consumption and cloud infrastructure billing rates when the devices and the cloud computing system adapt their resource consumption according to the volume of generated queries by switching between ``idle'' and ``active'' states.  Experiments with Beaglebone Linux embedded platforms and Amazon Web Services (AWS)\ based back-end processing for visual query generation, transmission and similarity detection demonstrate that the proposed model forms a framework that accurately incorporates the effect of various system parameters with respect to energy consumption and cloud billing costs. Therefore, variations of the proposed analytic modeling can be used for early-stage analysis of possible deployments, or limit studies of the expected performance under a wide range of parameter settings, prior to costly deployments in the field. Our framework could be expanded in future work by:\ \textit{(i)} expanding our analytic results beyond the specific cases of distributions used to characterize the query data volumes; \textit{(ii)} considering the case of simple aggregation of  the IoT\ devices' traffic by the IoT aggregator (Fig. \ref{fig:visual-search-system-diagram}); \textit{(iii)} extending the experimental validation to different testbeds and applications, e.g., within IBM\ IoT Foundation and Bluemix, AWS\ IoT, Cisco OpenStack, etc.

\section*{Acknowledgements}
FR, JD, and YA were supported by EPSRC, grants EP/K033166/1, EP/M00113X/1. VG was supported by  Innovate UK, project ACAME, grant no. 131983.

 \section*{Appendix}

 \subsection{Energy Consumption: $P(\psi_{\mathrm{e}})$ Is Pareto Distributed}
 
 In this case, $\Psi_{\mathrm{e}}$ is drawn from a Pareto distribution with scale $v_{\mathrm{e}}$ and shape $\alpha_{\mathrm{e}}$, with $\alpha_{\mathrm{e}}>2$, as in \eqref{eq:P_P_psi_e_Pareto_1}.
 Note that the expected value of $\Psi_{\mathrm{e}}$ is given by $E_\mathrm{P}[\Psi_{\mathrm{e}}]= \frac{\alpha_{\mathrm{e}} v_{\mathrm{e}}}{\alpha_{\mathrm{e}-1}}$. Therefore, we set $v_{\mathrm{e}} = \frac{\alpha_\mathrm{e}-1}{\alpha_{\mathrm{e}}} r$  in order to be consistent with the analysis carried out for the case of uniformly distributed $\Psi_\mathrm{e}$.
 
 If $c_{\mathrm{e}} r \leq v_{\mathrm{e}}$, the device is never in idle state, and the corresponding expected energy consumption is simply  $E_{\mathrm{exp,P}} = g_{\mathrm{e}} r$. Under $c_{\mathrm{e}} r > v_{\mathrm{e}}$, and by using \eqref{eq:P_P_psi_e_Pareto_1} in \eqref{eq:E_exp}, we obtain
 \begin{IEEEeqnarray}{rCl}
 E_{\mathrm{exp,P}} & =&  \left(  g_{\mathrm{e}}  + i_{\mathrm{e}}  \left(  (\alpha_{\mathrm{e}}-1)^{\alpha_{\mathrm{e}}-1} c_{\mathrm{e}} (\alpha_{\mathrm{e}} c_{\mathrm{e}})^{-\alpha_{\mathrm{e}}} + c_{\mathrm{e}}-1  \right)    \right)r. \IEEEeqnarraynumspace
  \label{eq:E_exp_P_r}
 \end{IEEEeqnarray}
 Thus, the value of the average query volume that meets the expected energy consumption constraint is
 \begin{equation}
 r = \frac{E_{\mathrm{exp,P}}}{ g_{\mathrm{e}}  + i_{\mathrm{e}}  \left(  (\alpha_{\mathrm{e}}-1)^{\alpha_{\mathrm{e}}-1} c_{\mathrm{e}} (\alpha_{\mathrm{e}} c_{\mathrm{e}})^{-\alpha_{\mathrm{e}}} + c_{\mathrm{e}}-1  \right)   }.
 \label{eq:r_E_exp_P}
 \end{equation}
 Similarly, by using \eqref{eq:P_P_psi_e_Pareto_1} in \eqref{eq:E_var}, and under $c_{\mathrm{e}} r > v_{\mathrm{e}}$, we can write the upper-sided variability of the energy consumption when the application switches from ``idle'' to ``active'' state as
 \begin{IEEEeqnarray}{rCl}
 E_{\mathrm{var,P}} & = & 2 g_{\mathrm{e}}^2 \frac{ (\alpha_{\mathrm{e}}-1)^{\alpha_{\mathrm{e}}-1}c_{\mathrm{e}}^{2 - \alpha_{\mathrm{e}}}  }{\alpha_{\mathrm{e}}^{\alpha_{\mathrm{e}}}(\alpha_{\mathrm{e}}-2)} r^{2}.
 \label{eq:E_var_P_r}
 \end{IEEEeqnarray}
 Then, by substituting \eqref{eq:r_E_exp_P} into \eqref{eq:E_var_P_r}, we can express the one-side variability of the energy consumption as
 \begin{IEEEeqnarray}{rCl}
 \nonumber
 E_{\mathrm{var,P}} & = &  g_{\mathrm{e}}^2 \frac{ (\alpha_{\mathrm{e}}-1)^{\alpha_{\mathrm{e}}-1}c_{\mathrm{e}}^{2 - \alpha_{\mathrm{e}}}  }{\alpha_{\mathrm{e}}^{\alpha_{\mathrm{e}}}(\alpha_{\mathrm{e}}-2)} \\
 & & \cdot \frac{  E_{\mathrm{exp,P}}^{2}  }{  \left(  g_{\mathrm{e}}  + i_{\mathrm{e}}  \left(  (\alpha_{\mathrm{e}}-1)^{\alpha_{\mathrm{e}}-1} c_{\mathrm{e}} (\alpha_{\mathrm{e}} c_{\mathrm{e}})^{-\alpha_{\mathrm{e}}} + c_{\mathrm{e}}-1  \right)    \right)^{2}}. \IEEEeqnarraynumspace
 \end{IEEEeqnarray}
 
 We note that $\alpha_{\mathrm{e}}>2$ is a necessary and sufficient condition for the upper-sided energy variability to be finite.

 \subsection{Energy Consumption: $P(\psi_{\mathrm{e}})$ Is Exponentially Distributed}
 
 Consider now the case where $\Psi_{\mathrm{e}}$ is exponentially distributed with rate parameter $\frac{1}{r}$ as in \eqref{eq:P_P_psi_e_Exponential_1},
 with the expected value of  $\Psi_{\mathrm{e}}$ set to $E_\mathrm{E}[\Psi_{\mathrm{e}}] = r$.
 
 The expected energy consumption at each device can be computed by substituting \eqref{eq:P_P_psi_e_Exponential_1} in \eqref{eq:E_exp}, thus obtaining
 \begin{equation}
 E_{\mathrm{exp,E}} = \left(  g_{\mathrm{e}}  + i_{\mathrm{e}} \left(c_{\mathrm{e}} + e^{-c_{\mathrm{e}} } -1  \right)  \right) r.
 \label{eq:E_exp_exponential}
 \end{equation}
 Moreover, the one-side variability for the energy consumption when the application switches from idle to active state is obtained by substituting \eqref{eq:P_P_psi_e_Exponential_1} in \eqref{eq:E_var}, leading to
 \begin{equation}
 E_{\mathrm{var,E}} = 2 g_{\mathrm{e}}^2 e^{-c_{\mathrm{e}}} {r^2} .
 \label{eq:E_var_exponential}
 \end{equation}
 From \eqref{eq:E_exp_exponential}, we can derive the average query volume that meets the average energy consumption constraint in function of the activation rate $c_{\mathrm{e}}$:
 \begin{equation}
 r = \frac{E_{\mathrm{exp,E}}}{g_{\mathrm{e}}  + i_{\mathrm{e}} \left(c_{\mathrm{e}} + e^{-c_{\mathrm{e}} } -1  \right) }.
 \label{eq:r_E_exp_exponential}
 \end{equation}
 Then, by substituting \eqref{eq:r_E_exp_exponential} in \eqref{eq:E_var_exponential}, we obtain the expression of the upper-sided energy variability associated to a single device as a function of the activation rate $c_{\mathrm{e}}$  
 \begin{equation}
 E_{\mathrm{var,E}} = \frac{  2 g_{\mathrm{e}}^2 e^{-c_{\mathrm{e}}}  E_{\mathrm{exp,E}}^2}{\left(  g_{\mathrm{e}}  + i_{\mathrm{e}} \left(c_{\mathrm{e}} + e^{-c_{\mathrm{e}} } -1  \right)  \right)^2}.
 \end{equation}
 
 Provided that $E_{\mathrm{exp,E}} > g_{\mathrm{e}} r$, we can also determine the activation rate $c_{\mathrm{e}}$ that guarantees a given average energy consumption constraint for any average query volume. This is achieved by solving equation \eqref{eq:E_exp_exponential} for $c_{\mathrm{e}}$, thus obtaining
 \begin{equation}
 c_{\mathrm{e}} = W_0 \left( - \exp\left(  -  \frac{  E_{\mathrm{exp,E}} + i_{\mathrm{e}}r - g_{\mathrm{e}}r  }{i_{\mathrm{e}}r}   \right)      \right) + \frac{  E_{\mathrm{exp,E}} + i_{\mathrm{e}}r - g_{\mathrm{e}}r  }{i_{\mathrm{e}}r},
 \end{equation}
 where $W_0(\cdot)$ is the main branch of the standard Lambert W function \cite{corless1996lambertw}. Finally, the corresponding one-side energy variability is given by
 \begin{equation}
 E_{\mathrm{var,E}}  = - 2 g_{\mathrm{e}}^2 r^2 W_0 \left( - \exp\left(  -  \frac{  E_{\mathrm{exp,E}} + i_{\mathrm{e}}r - g_{\mathrm{e}}r  }{i_{\mathrm{e}}r}   \right)      \right).
 \end{equation}

\bibliographystyle{IEEEtran}
\bibliography{literatur}

\end{document}